# Lattice Boltzmann modeling of self-propelled Leidenfrost droplets on ratchet surfaces

Qing Li [a, b], Q. J. Kang [b, *], M. M. Francois [c], and A. J. Hu [d,]

[a] School of Energy Science and Engineering, Central South University, Changsha 410083, China

[b] Computational Earth Science Group, Los Alamos National Laboratory, Los Alamos, NM 87545, USA

[c] Fluid Dynamics and Solid Mechanics, Los Alamos National Laboratory, Los Alamos, NM 87545, USA

[d] School of Civil Engineering and Architecture, Southwest University of Science and Technology, Mianyang 621010, China

**Abstract**

In this paper, the self-propelled motion of Leidenfrost droplets on ratchet surfaces is numerically investigated with a thermal multiphase lattice Boltzmann model with liquid-vapor phase change. The capability of the model for simulating evaporation is validated via the $D^2$ law. Using the model, we first study the performances of Leidenfrost droplets on horizontal ratchet surfaces. It is numerically shown that the motion of self-propelled Leidenfrost droplets on ratchet surfaces is owing to the asymmetry of the ratchets and the vapor flows beneath the droplets. It is found that the Leidenfrost droplets move in the direction toward the slowly inclined side from the ratchet peaks, which agrees with the direction of droplet motion in experiments [Linke *et al.*, *Phys. Rev. Lett.*, 2006, **96**, 154502]. Moreover, the influences of the ratchet aspect ratio are investigated. For the considered ratchet surfaces, a critical value of the ratchet aspect ratio is approximately found, which corresponds to the maximum droplet moving velocity. Furthermore, the processes that the Leidenfrost droplets climb uphill on inclined ratchet surfaces are also studied. Numerical results show that the maximum inclination angle at which a Leidenfrost droplet can still climb uphill successfully is affected by the initial radius of the droplet.

*Corresponding author: qkang@lanl.gov

## 1. Introduction

When a liquid droplet is deposited on a solid surface whose temperature is far above the boiling point of the liquid (e.g., the room temperature for liquid nitrogen), the droplet will be levitated above the hot surface through the action of a vapor layer below its bottom surface [1, 2]. The vapor layer prevents the droplet from contacting the hot surface, which considerably reduces the heat transfer and therefore retards the evaporation of the droplet [3, 4]. This phenomenon, known as the "Leidenfrost phenomenon" [5, 6], has attracted significant attention due to the fact that the Leidenfrost state is a perfect superhydrophobic state [3] and can provide an almost frictionless motion [1].

The liquid droplets at the Leidenfrost state are usually called *Leidenfrost droplets*. The characteristic of the vapor layer beneath a Leidenfrost droplet has been studied by Biance *et al*. [7]. In addition, they have deduced the scaling laws for Leidenfrost droplets. In 2006, Linke *et al*. [8] found that the Leidenfrost droplets perform self-propelled motion when they are placed on hot surfaces with asymmetric textures. They showed that [8] a solid surface covered with asymmetric ratchets and heated over the Leidenfrost temperature (the minimum temperature for stable film boiling) is able to propel an evaporating droplet in a preferential direction. This discovery has been recognized as a key modern breakthrough in controlling Leidenfrost droplets [1, 2]. It is believed that the feature of self-propelled Leidenfrost droplets can be utilized to create devices in which self-propulsion is obtained [1].

Following the work of Linke *et al*., many experimental studies have been conducted in recent years about the Leidenfrost phenomena. In 2011, Ok *et al*. [9] studied Leidenfrost droplets on micro- and submicron-ratchet surfaces and showed that a hydrophobic coating on ratchet surfaces can increase the velocity of a Leidenfrost droplet and decrease the threshold temperature of the droplet motion. Lagubeau *et al*. [10] found that the Leidenfrost solids (such as dry ice) also self-propel on hot ratchets in the same direction as liquids. Nevertheless, they proposed that the Leidenfrost droplets are driven by an inertial propelling force. Such an issue was later clarified by Dupeux *et al*. [11] and Baier *et al*. [12]. They demonstrated that the viscous vapor flow is the main propulsion mechanism for Leidenfrost droplets, which is usually called the "viscous mechanism" in comparison with the inertial mechanism proposed in Ref. [10]. Furthermore, Marin *et al*. [13, 14] have studied Leidenfrost droplets on micro-ratchets with different droplet initial sizes, ratchet geometries and temperatures. They found that the viscous mechanism fits reasonably well with the experiments performed on micro-ratchets, for both large droplets and capillary droplets.

Cousins *et al*. [2] have reported a circular ratchet trap (a surface with concentric circular ridges, each asymmetric in cross-section) for Leidenfrost droplets and Feng *et al*. [15] reported a ratchet composite thin film for low-temperature self-propelled Leidenfrost droplets. In addition, Dupeux *et al*. [16] have studied the effects of crenelated surfaces on sliding Leidenfrost droplets and Würger [17] has investigated the thermal creep in the motion of self-propelled Leidenfrost droplets/solids. However, Würger claimed that the thermal creep flow could be the origin of the propulsion mechanism. Later, Hardt *et al*. [18] examined the nature of thermally driven flows and quantified their contribution to the propulsion, showing that thermally driven flows make an insignificant contribution to the thrust of Leidenfrost objects. Moreover, Grounds *et al*. [19] have experimentally studied the processes that Leidenfrost droplets climb uphill on tilted ratchet surfaces with different sub-structures. Recently, Dupeux *et al*. [20] reported that the water droplets can be propelled far below the usual Leidenfrost temperature when using textured superhydrophobic ratchets, which extends the parameter range where self-propulsion can be obtained. Using a low pressure environment, Celestini *et al*. [21] found that the Leidenfrost droplets of water can be generated at room temperature. In addition, Celestini and Kirstetter [22] have investigated the influence of an electric field on Leidenfrost droplets and Maquet *et al*. [23] have studied the organization of microbeads in Leidenfrost droplets. Most recently, Wells *et al*.

[24] designed a Leidenfrost-based engine, which converts temperature difference into mechanical work through the Leidenfrost effect on turbine-like surfaces. It has been widely found that the Leidenfrost objects (droplets and solids) on ratchet surfaces move toward the slowly inclined side from the ratchet peaks. Nevertheless, anomalous cases have also been observed by Ok [25] and Hashmi *et al*. [26] on ratchet surfaces with metal oxide and chemically-contaminated ratchet surfaces, respectively.

With the rapid development of computational science and computer hardware, numerical simulation gradually plays an important role in scientific research. In many research fields, numerical simulation has become an alternative to experiments or serves as an important supplement to experimental studies. About the viscous vapor flow between a Leidenfrost droplet and a ratchet surface, some theoretical explanations/assumptions have been made in the literature [8, 11-13]. However, to date, there is still no direct information about the vapor flow below a self-propelled Leidenfrost droplet. Definitely, the detailed flow field provided by numerical simulations would be very useful for revealing or demonstrating the fundamental features of self-propelled Leidenfrost droplets.

To the best of our knowledge, there have been very few numerical studies that are related to self-propelled Leidenfrost droplets, which may be attributed to the challenge of modeling interfacial dynamics on rough surfaces with phase-change heat transfer. The existing numerical studies of Leidenfrost droplets were mostly focused on the shape of Leidenfrost droplets on a flat surface [27-29] and the impact of droplets on flat hot surfaces in the Leidenfrost regime [30, 31], e.g., Xu and Qian [28] and Bouwhuis *et al*. [29] have recently conducted numerical simulations of Leidenfrost droplets on a flat surface. For a Leidenfrost droplet on a flat horizontal surface, the expanding vapor will flow outward equally in all directions, which will not lead to self-propelled motion in a preferred direction. It is also noticed that Cousins *et al*. [2] have simulated airflow over a ratchet, but without considering the thermal processes involved in evaporation and levitation.

The purpose of the present work is to investigate the self-propelled Leidenfrost droplets from the numerical point of view. Specifically, the lattice Boltzmann method [32-37], which can be viewed as a discrete solver for the Boltzmann equation in the kinetic theory, is employed to simulate the dynamics of Leidenfrost droplets on ratchet surfaces. This method has been applied in a variety of fields with great success [38] and has been recently utilized to simulate liquid-vapor phase change, such as boiling heat transfer [39, 40] and droplet evaporation [41]. The pseudopotential multiphase lattice Boltzmann model is adopted [36, 37], which is very popular in the lattice Boltzmann community because of its distinct advantage. In this model, the phase separation between different phases can emerge automatically as a result of particle interactions [42, 43], without the need to use any technique to track or capture the liquid-vapor interface. The rest of the present paper is organized as follows. The adopted multiphase lattice Boltzmann model is introduced in Section 2. The numerical simulations of self-propelled Leidenfrost droplets and the discussion are presented in Section 3. A brief summary is finally given in Section 4.

## 2. Model description

In the past two decades, the lattice Boltzmann method has been developed into an efficient mesoscopic numerical method for simulating fluid flows and heat transfer [33-35]. Unlike traditional numerical methods that simulate fluid flows by directly solving the Navier-Stokes equations, the lattice Boltzmann method is based on the mesoscopic kinetic equation. It simulates fluid flows by solving the discrete Boltzmann equation with certain collision operators, such as the Bhatnagar-Gross-Krook collision operator [44] and the Multiple-Relaxation-Time (MRT) collision operator [45, 46], and then accumulating the density distribution function to obtain the macroscopic averaged properties [33]. Using the Chapman-Enskog analysis, it can be found that the macroscopic Navier-Stokes equations can be recovered from the lattice Boltzmann equation [33].

Generally, the lattice Boltzmann equation, which governs the evolution of the density distribution function, can be written as follows:

$$f_\alpha(\mathbf{x}+\mathbf{e}_\alpha\delta_t, t+\delta_t) - f_\alpha(\mathbf{x},t) = \Omega_\alpha(\mathbf{x},t) + \delta_t F'_\alpha(\mathbf{x},t), \qquad (1)$$

where $f_\alpha$ is the density distribution function, $\mathbf{x}$ is the spatial position, $\mathbf{e}_\alpha$ is the discrete velocity along the $\alpha$ th lattice direction, $\delta_t$ is the time step, $\Omega_\alpha$ is the collision term, and $F'_\alpha$ is the forcing term. The MRT collision operator is adopted, which can be written as $\Omega_\alpha = -(\mathbf{M}^{-1}\mathbf{\Lambda}\mathbf{M})_{\alpha\beta}(f_\beta - f_\beta^{eq})$ [45, 46], where $\mathbf{M}$ is an orthogonal transformation matrix, $\mathbf{\Lambda}$ is a diagonal Matrix, and $f_\beta^{eq}$ is the equilibrium distribution. The detailed forms of $\mathbf{M}$ and $\mathbf{\Lambda}$ can be found in Ref. [45]. Using the transformation matrix $\mathbf{M}$, the right-hand side of Eq. (1) can be rewritten as

$$\mathbf{m}^* = \mathbf{m} - \mathbf{\Lambda}(\mathbf{m} - \mathbf{m}^{eq}) + \delta_t\left(\mathbf{I} - \frac{\mathbf{\Lambda}}{2}\right)\mathbf{S} \qquad (2)$$

where $\mathbf{m} = \mathbf{M}\mathbf{f}$, $\mathbf{m}^{eq} = \mathbf{M}\mathbf{f}^{eq}$, $\mathbf{I}$ is the unit tensor, and $\mathbf{S}$ is the forcing term in the moment space. The detailed form of $\mathbf{S}$ can be found in Refs. [40, 47]. Then the lattice Boltzmann equation is

$$f_\alpha(\mathbf{x}+\mathbf{e}_\alpha\delta_t, t+\delta_t) = f_\alpha^*(\mathbf{x},t), \qquad (3)$$

where $\mathbf{f}^* = \mathbf{M}^{-1}\mathbf{m}^*$. The macroscopic density and velocity are calculated via

$$\rho = \sum_\alpha f_\alpha, \quad \rho\mathbf{v} = \sum_\alpha \mathbf{e}_\alpha f_\alpha + \frac{\delta_t}{2}\mathbf{F}, \qquad (4)$$

where $\mathbf{F} = (F_x, F_y)$ is the total force acting on the system.

The gravitational force is given by $\mathbf{F}_b = (\rho - \rho_V)\mathbf{g}$, where $\rho_V$ is the vapor-phase density and $\mathbf{g} = (0, -g)$ is the gravitational acceleration. For single-component multiphase systems, the intermolecular interaction force $\mathbf{F}_m$ [36, 37], through which the phase separation between difference phases can be automatically achieved, is given as follows [48]:

$$\mathbf{F}_m = -G\psi(\mathbf{x})\sum_\alpha w_\alpha \psi(\mathbf{x}+\mathbf{e}_\alpha\delta_t)\mathbf{e}_\alpha \qquad (5)$$

where $\psi(\mathbf{x})$ is the pseudopotential, $G$ is the interaction strength, and $w_\alpha$ are the weights [48]. To reproduce a non-ideal equation of state, the pseudopotential is taken as $\psi(\mathbf{x}) = \sqrt{2(p_{\text{EOS}} - \rho c_s^2)/Gc^2}$, where $p_{EOS}$ is the non-ideal equation of state. In the original pseudopotential lattice Boltzmann model devised by Shan and Chen [36, 37], the surface tension cannot be tuned independently of the liquid-vapor density ratio. Using the treatment in Ref. [49], which was proposed to decouple the surface tension from the density ratio, the surface tension can be an adjustable parameter when the density ratio is fixed.

Through the Chapman-Enskog analysis, it can be found that the Navier-Stokes equation with a non-ideal pressure tensor can be recovered from Eqs. (2) and (3) [40]. The governing equation for the temperature field is given by (the viscous heat dissipation is neglected) [50]

$$\rho c_v \frac{DT}{Dt} = \nabla \cdot (\lambda \nabla T) - T \left( \frac{\partial p_{EOS}}{\partial T} \right)_\rho \nabla \cdot \mathbf{v}, \qquad (6)$$

where $\lambda$ is the thermal conductivity and $c_v$ is the specific heat at constant volume. The temperature equation is solved with the fourth-order Runge-Kutta scheme for time discretization and the isotropic central scheme for spatial discretization [40]. The Peng-Robinson equation of state is adopted [51]

$$p_{EOS} = \frac{\rho RT}{1-b\rho} - \frac{a\varphi(T)\rho^2}{1+2b\rho - b^2\rho^2}, \qquad (7)$$

where $\varphi = \left[1 + \left(0.37464 + 1.54226\omega - 0.26992\omega^2\right)\left(1 - \sqrt{T/T_c}\right)\right]^2$ ($\omega = 0.344$), $a = 0.45724 R^2 T_c^2 / p_c$, and $b = 0.0778 RT_c / p_c$. The parameters $a$, $b$, and $R$ are chosen as $a = 3/49$, $b = 2/21$, and $R = 1$ [40]. The critical temperature $T_c$ can be obtained from the formulations of $a$ and $b$. Note that all the quantities in the present paper are taken in lattice units, namely the units in the lattice Boltzmann method, which are based on the lattice constant $c = \delta_x / \delta_t = 1$, where $\delta_x$ is the spatial spacing and $\delta_t$ is the time step. The conversion between the lattice units and the physical units can be found, e.g., in Refs. [34, 52]. Obviously, using a non-ideal equation of state, the pseudopotential $\psi$ in Eq. (5) will be linked to the temperature field. Then the liquid-vapor phase change can be driven by the temperature field through the equation of state. As a result, the rate of the liquid-vapor phase change is a computational output [28] rather than an artificial input, which is implemented by adding artificial phase-change terms to the temperature equation [53]. Biferale *et al*. [39] have numerically demonstrated that the Clausius-Clapeyron relation is satisfied when using a non-ideal equation of state for simulating liquid-vapor phase change. Nevertheless, such a treatment for thermal multiphase flows is currently only applicable to low and moderate liquid-vapor density ratios due to the problem that at large density ratios the interface thickness considerably changes with the temperature [47]. Therefore in the present work the saturation temperature is chosen to be $T_{\text{sat}} = 0.86T_c$, which corresponds to the liquid-vapor density ratio $\rho_L/\rho_V \approx 17$.

## 3. Numerical results and discussion

### 3.1 Validation of the D² law.

In this section numerical simulations are conducted to assess the validity of the model for simulating evaporation. The well-know D² law for droplet vaporization is considered [54, 55], which predicts the time rate of change of the square of the evaporating droplet diameter to be constant, i.e., $D^2(t) = D_0^2 - Kt$, on the basis of the following conditions: the liquid and vapor phases are quasi-steady, the evaporation occurs in an environment with negligible viscous heat dissipation and no buoyancy ($g = 0$), and the thermophysical properties ($c_p$, $c_v$, and $\lambda$) are constant. Our simulations are carried out on a computational domain (an enclosed cavity) discretized by $N_x \times N_y = 200 \times 200$ lattices nodes with a droplet (the droplet diameter $D_0 = 60$) being initially placed at the center. The temperature of the droplet is the saturation temperature $T_{\text{sat}}$.

At the initial time step, a uniform temperature $T_g$ is applied to the surrounding vapor of the droplet. The superheat $\Delta T = T_g - T_{\text{sat}}$ is chosen to be $0.14T_c$. The evaporation is caused by the temperature gradient at the liquid-vapor interface. During the process, the vapor phase temperature is kept above the droplet temperature by employing a constant temperature condition ($T_g$) at the boundaries. The kinematic viscosity is set to $v = 0.1$ ($\tau_v = 0.8$) in the whole computational domain. The specific heat at constant volume $c_v$ is taken as $c_v = 5$. According to the assumptions for the $D^2$ law, the thermal conductivity $\lambda$ should be constant. Then the parameter $K$ in the $D^2$ law will depend linearly on $\lambda$ [54]. Two cases are considered, i.e., Case A: $\lambda = 1/3$ and Case B: $\lambda = 2/3$ (lattice unit). The evaporation processes of these two cases are displayed in Fig. 1.

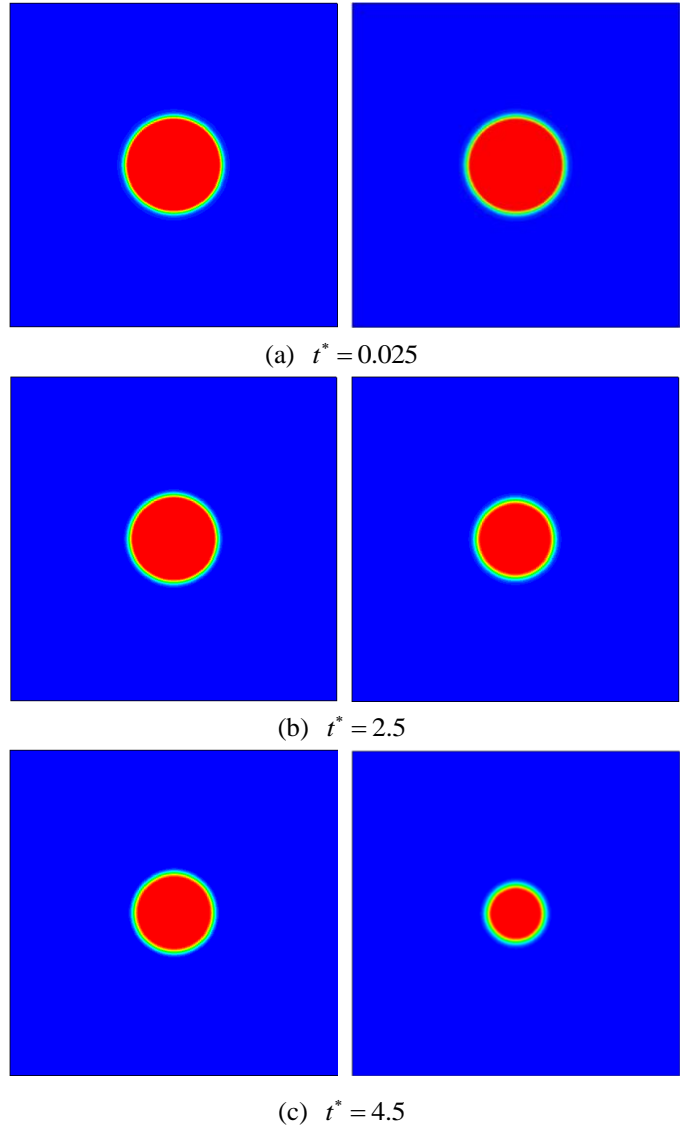

(a) $t^* = 0.025$

(b) $t^* = 2.5$

(c) $t^* = 4.5$

**Fig. 1** Snapshots of droplet evaporation at $t^* = 0.025$, $2.5$, and $4.5$. Case A (left) and Case B (right).

The non-dimensional time $t^*$ is defined as $t^* = t/t_h$, where $t_h = \rho l_M^2 / \mu = l_M^2 / v$ is the minimum hydrodynamic time scale [34] (see Section 1.4 in the reference), in which $l_M$ is a typical length scale, $\mu = \rho \nu$ is the dynamic viscosity, and $\nu$ is the kinematic viscosity. For the present problem, the characteristic length scale, $l_M$, is chosen to be the droplet diameter $l_M = D_0$. A similar characteristic time can also be found in Ref. [56]. From Fig. 1 it can be seen that for both cases the circular shape of the droplet is well preserved during the evaporation process. Furthermore, we can see that the droplet evaporates faster in Case B than in Case A. The square of the non-dimensional droplet diameter ($D/D_0$) is plotted against time for the two cases in Fig. 2. From the figure, the $D^2$ law, namely the linear relationship between $(D/D_0)^2$ and $t^*$, can be clearly observed. Moreover, the evaporation rates (the slope denoted by the parameter $K$ in the $D^2$ law) of Cases A and B are found to be $K \approx 0.0037$ and $0.0076$, respectively. It can be seen that the evaporation rate of Case B is approximately two times that of Case A.

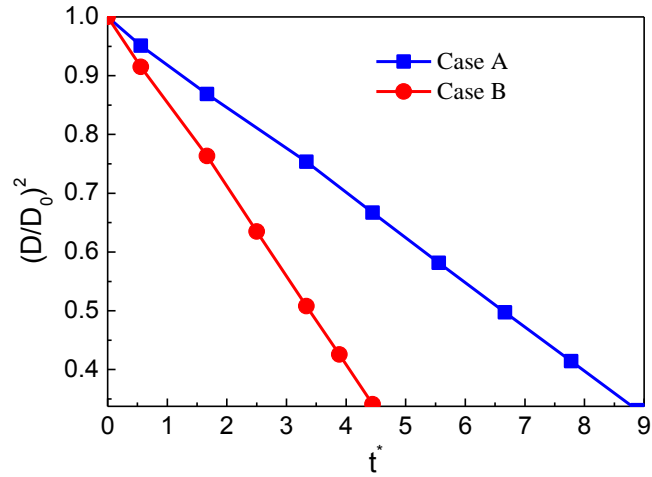

**Fig. 2** The variation of the square of the non-dimensional droplet diameter with time.

### 3.2 Simulation of Leidenfrost droplets on ratchet surfaces

3.2.1 Numerical setup

Numerical simulations are now carried out to investigate the behavior of Leidenfrost droplets on ratchet surfaces. The geometric structure of the ratchet surface is illustrated in Fig. 3, which has been adopted in several experimental studies. In the figure, $L$ and $H$ are defined according to Ref. [9]

(see Figs. 1 and 3(a) in the reference). It can be found that $H=0$ and $H=L$ correspond to flat surfaces and symmetric ratchet surfaces, respectively. For such a structure, we adopt the standard Cartesian coordinate system with the *x* and *y* directions being parallel to the directions of $L$ and $H$, respectively. For horizontal ratchet surfaces, the gravitational force is perpendicular to the dash-dotted line in Fig. 3.

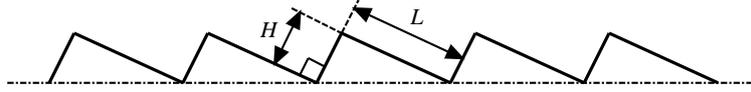

**Fig. 3** A schematic diagram of the ratchet surface.

The temperature of the ratchet surface is $T_w$ and the surface superheat $\Delta T = T_w - T_{sat}$ is fixed at $0.34T_c$, which is sufficient for generating the Leidenfrost phenomenon. The relaxation time $\tau_v$ is set to 0.9 (the kinematic viscosity $v = (\tau_v - 0.5)/3$). The thermal conductivity $\lambda = \rho c_v \chi$ is chosen to be proportional to the density $\rho$ with a constant $\chi$, which is the thermal diffusivity [40]. The ratio between the thermal diffusivity $\chi$ and the kinematic viscosity $v$ is taken as $\chi/v = 0.35$. The gravitational acceleration $g$ is set to $g = 10^{-5}$. As previously mentioned, all the quantities are taken in lattice units and can be directly implemented in the numerical codes.

Using numerical simulations, there are many parameters that can be investigated. Nevertheless, as a preliminary attempt in modeling self-propelled Leidenfrost droplets on ratchet surfaces, we focus on investigating the performances of Leidenfrost droplets with different initial radii and the effects of the aspect ratio ($H/L$) of the ratchet. Moreover, we will study the processes that the Leidenfrost droplets climb uphill on tilted ratchet surfaces. In our simulations, $L$ is fixed at $L = 48$ (lattice unit). Five different choices of $H$ are considered when studying the effects of $H$ ($H/L = 1/4$, $7/24$, $1/3$, $5/12$, and $1/2$). For the former three cases, the grid system is $N_x \times N_y = 500 \times 250$. For the latter two cases, the grid systems are $500\times300$ and $500\times360$, respectively. The no-slip boundary condition is applied to the ratchet surface and the out flow boundary condition is employed at non-wall boundaries. The unknown information at the out flow boundary is extrapolated from the interior field.

3.2.2 The vapor flow and the effects of the initial droplet radius

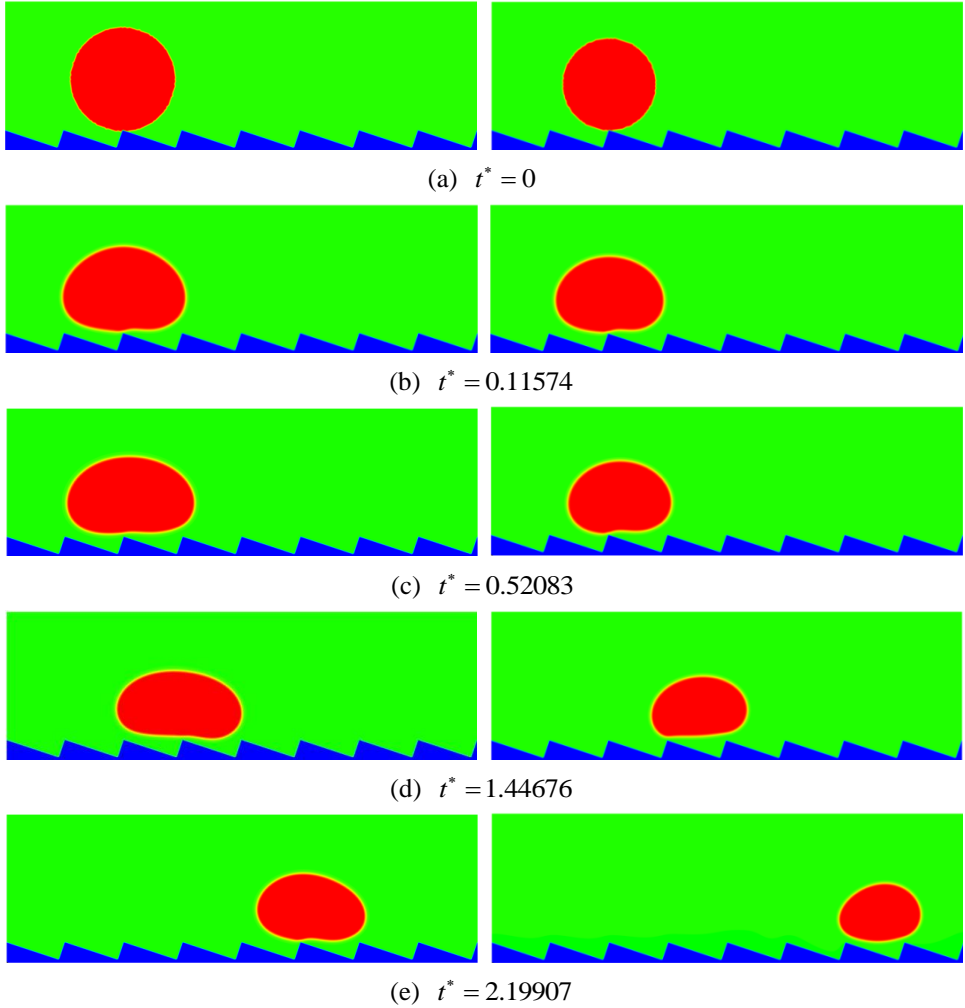

(a) $t^* = 0$

(b) $t^* = 0.11574$

(c) $t^* = 0.52083$

(d) $t^* = 1.44676$

(e) $t^* = 2.19907$

**Fig. 4** Snapshots of the motion of Leidenfrost droplets on a horizontal ratchet surface with $H/L=1/3$. The initial droplet radius is $R_0 = 45$ (left) and $R_0 = 40$ (right).

In the present study, the initial radius ($R_0$) of the droplet is considered to be comparable with $L$. In this subsection we report the simulations of Leidenfrost droplets on a horizontal ratchet surface with $H/L=1/3$. The non-dimensional time is also defined as $t^* = t/t_h$ with $t_h = l_M^2/\nu$ (see the descriptions in Section 3.1). Here $l_M$ is chosen to be $L$. The results of the cases $R_0 = 45$ and 40 are shown in Fig. 4. It should be noted that the computational domain is larger than the domain shown in Fig. 4 (as well as the following figures). As can be seen in Fig. 4(a), the droplets were initially placed in contact with the peak of a ratchet. After the initial time step, the liquids near the contact point will evaporate very rapidly because of the high temperature of the ratchet surface. The released vapor will form a thin vapor layer beneath the droplet, keeping the droplet away from the peak of the ratchet, which can be observed in Fig. 4(b). Moreover, from Fig. 4(b) we can see that the symmetry of the bottom surface of the droplet has been broken due to the asymmetry of the ratchet. Specifically, on the right-hand side of the ratchet peak, a portion of the droplet bottom surface has become concave.

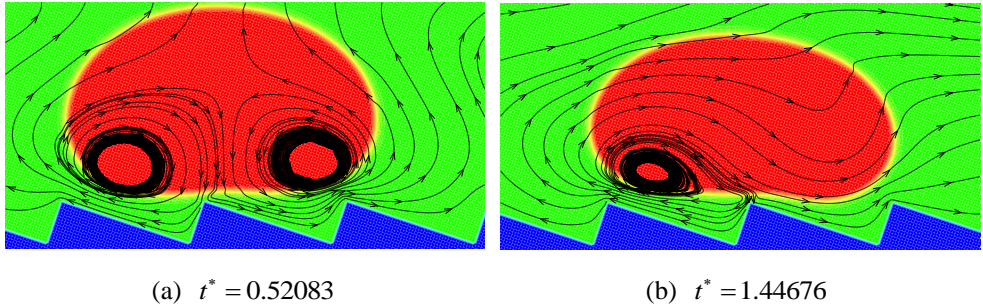

(a) $t^* = 0.52083$          (b) $t^* = 1.44676$

**Fig. 5** The streamlines of the case $R_0 = 45$ at $t^* = 0.52083$ and $1.44676$.

The streamlines of the case $R_0 = 45$ at $t^* = 0.52083$ and $1.44676$ are displayed in Fig. 5, which corresponds to the density contours in the left-hand panels of Figs. 4(c) and 4(d). From Fig. 5 the detailed flow information of the vapor layer between the droplet and the ratchet surface can be observed. Owing to the proximity of the ratchet peak to the droplet bottom surface, the evaporation in Fig. 5 is believed to be strongest around the peak of the central ratchet, which leads to very rapid generation of vapor. The evaporated vapor is then split by the ratchet peak and escapes along the trenches formed by the ratchets. To be specific, on the left-hand side of the central ratchet peak, the vapor flows toward the upper left. This flow supports the weight of the left part of the droplet and provides an "upward kick" to the droplet. Meanwhile, the vapor flow on the right-hand side pulls the droplet forward along the trench formed by the ratchets. As a result, the droplet moves to the right, which can be clearly seen in Fig. 4. In other words, it can be seen that the droplet moves in the direction toward the slowly inclined side from the ratchet peaks, which is consistent with the direction of droplet motion observed experimentally by Linke *et al*. [8].

From Fig. 5 we can also observe the flow information of the droplet. It can be found that the velocity vectors vary spatially within the droplet, which is attributed to the fact that the droplet is a soft deformable body rather than a solid object. It can be seen that the liquid circulations inside the droplet are affected by the droplet velocity. At $t^* = 0.52083$, the moving velocity of the droplet is very small. Therefore the liquid circulations inside the droplet are mainly induced by the vapor flows below the droplet, which leads to two vortices that rotate in opposite directions (one clockwise and the other counter-clockwise). At $t^* = 1.44676$, with the increase of the droplet velocity, the liquid circulations inside the droplet are changed. The counter-clockwise vortex has disappeared due to the significant increase of the droplet velocity and the fact that on the right-hand side of the central ratchet peak the droplet and the vapor move in the same direction. Meanwhile, the clockwise vortex, which was induced by the vapor flow on the left-hand side of the central ratchet peak, has become very small, resulting from the increasing influence of the droplet velocity.

The effects of the initial droplet radius are depicted in Fig. 6. Actually, in Fig. 4 it has been shown that the droplet with an initial radius of $R_0 = 40$ moves faster than that with $R_0 = 45$. To enable a more comprehensive comparison, the results of the case $R_0 = 35$ are also considered. The effect of the initial droplet radius on the average moving velocity of the droplet can be found in Fig. 6(a). The average droplet velocity (in lattice unit) at the time $t$ is measured via $\bar{u}(t) = \Delta s/\Delta t$, where $\Delta s$ is the distance (along the horizontal direction) traveled by the droplet between the time $t - \Delta t/2$ and $t+\Delta t/2$. From Fig. 6(a) we can see that the moving velocity of the droplet increases when the initial droplet radius decreases from $R_0 = 45$ to $R_0 = 35$. In our simulations, the surface tension $\gamma \approx 0.0852$, $\rho_L \approx 6.5$, and $g = 10^{-5}$ (lattice unit). Therefore the capillary length $l_{Ca} = \sqrt{\gamma/\rho_L g}$ is about $36.2$, which means that the radii of the droplets are comparable with the capillary length. In this regard, our numerical results are consistent with the experimental study of Marin *et al.* [13], who found that, when the droplet size is comparable with the capillary length, the droplet velocity will increase as the droplet size decreases.

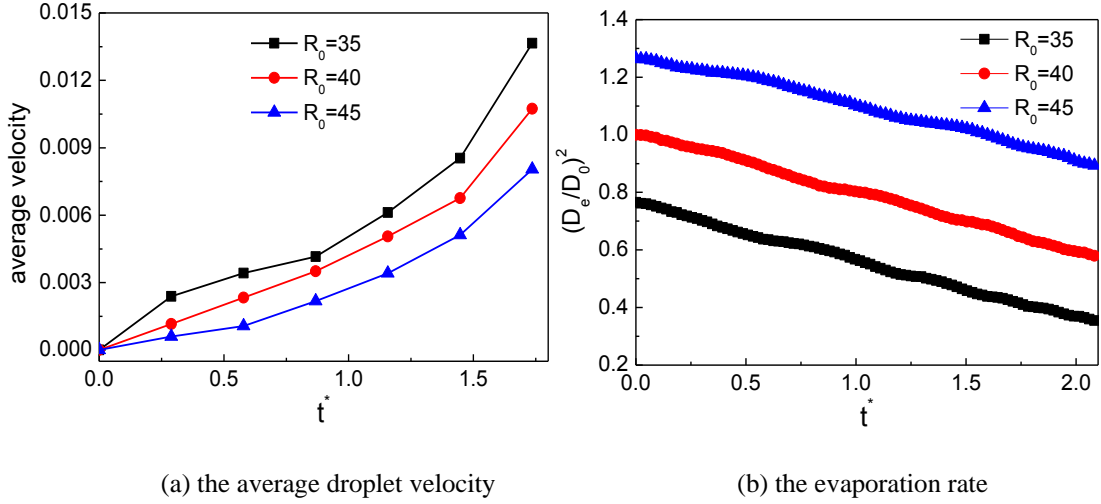

(a) the average droplet velocity  (b) the evaporation rate

**Fig. 6** The effects of the initial radius on the moving velocity of the droplet and the evaporation rate.

The evaporation rates of the three cases are also investigated and the results are shown in Fig. 6(b), where $D_e$ represents the effective diameter of the droplet, which is evaluated from the domain occupied by the droplet, and $D_0$ is the initial droplet diameter of the case $R_0 = 40$. From Fig. 6(b) we can find that there are no significant differences between the slopes of the three cases, which means that the amount of liquid that has been evaporated within a given time interval is approximately the same for the three cases, although the moving velocity of the droplet is different in these cases.

3.2.3 The influences of the ratchet aspect ratio

Now attention turns to the influences of the ratchet aspect ratio $H/L$. There are two special cases for this ratio: (a) the ratchet surface will reduce to a flat surface if $H$ equals zero ($H/L=0$) and (b) the ratchets will be symmetric when $H$ is equal to $L$ ($H/L=1$). Melling [57] has demonstrated that the Leidenfrost droplets placed on a horizontal flat surface or a symmetric ratchet surface at zero incline will not undergo directed motion. In other words, the self-propelled motion of Leidenfrost droplets in a preferential direction occurs when $H/L>0$ but disappears when $H/L$ is close to $1.0$. It is therefore believed that there probably exists a critical value of $H/L$ in the interval $[0,\ 1]$, which may provide the maximum propulsion for the Leidenfrost droplets.

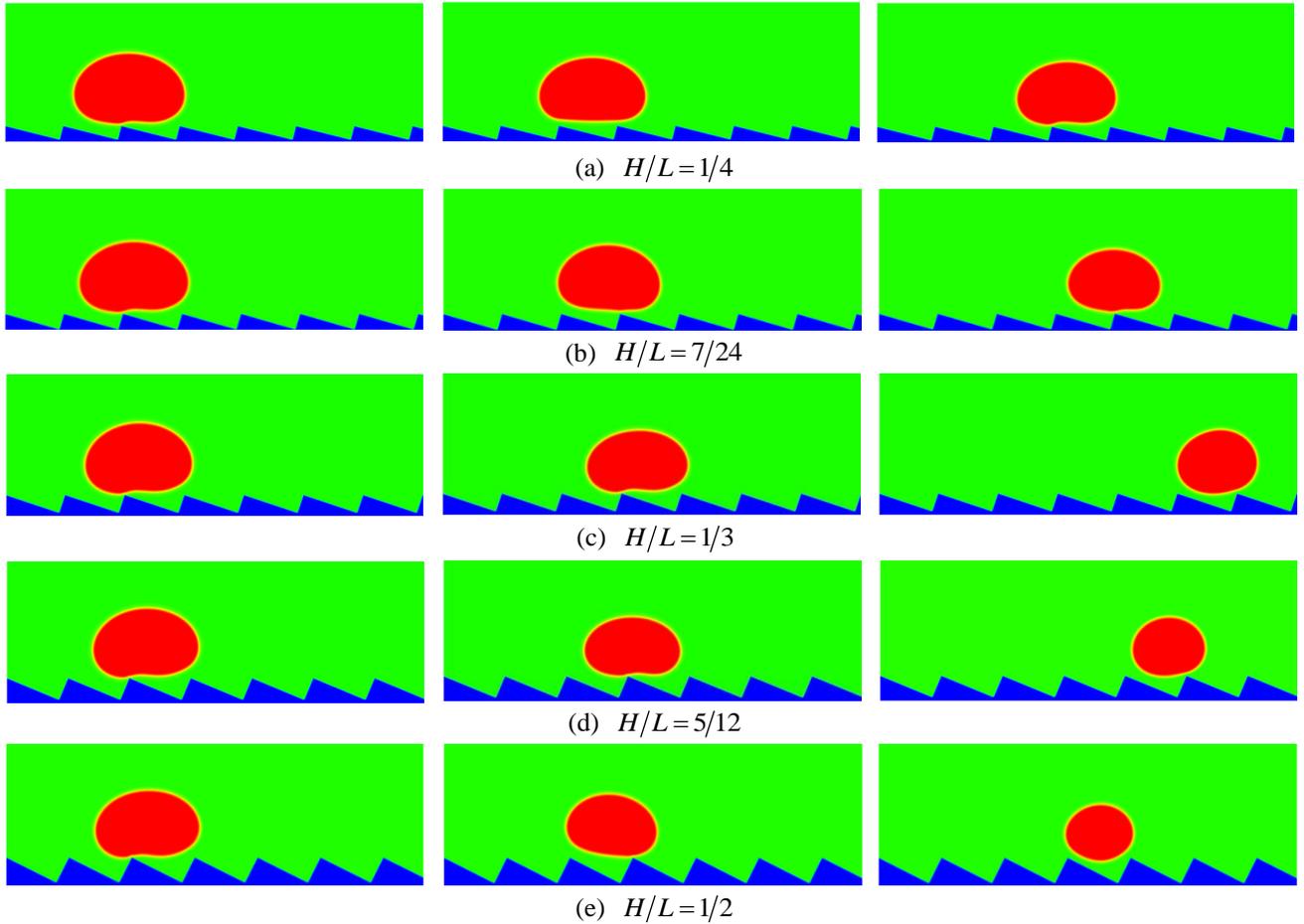

(a) $H/L=1/4$

(b) $H/L=7/24$

(c) $H/L=1/3$

(d) $H/L=5/12$

(e) $H/L=1/2$

**Fig. 7** Snapshots of the self-propelled motion of a Leidenfrost droplet with $R_0=40$ on five different ratchet surfaces at $t^*=0.57870$ (left), $1.33102$ (middle), and $2.02546$ (right).

To numerically evaluate the critical value of $H/L$, five different choices of $H$ are considered with $L$ being fixed, which have been previously mentioned: $H/L = 1/4$, $7/24$, $1/3$, $5/12$, and $1/2$. The self-propelled motion of a Leidenfrost droplet with $R_0 = 40$ on these ratchet surfaces is displayed in Fig. 7. The left, middle, and right panels of Fig. 7 represent the results at $t^* = 0.57870$, $1.33102$, and $2.02546$, respectively. By comparing the droplet positions in the three panels, we can see that the droplet moves faster when $H/L$ increases from $1/4$ to $1/3$. However, when $H/L$ is further increased, the moving velocity of the droplet decreases. To be specific, the left and right panels of Fig. 7 clearly show that, within the time interval $t^* \in [0.57870, 2.02546]$, the droplet traveled a much longer distance in the case $H/L = 1/3$ than in the other cases. Quantitatively, the average moving velocity of the droplet is plotted in Fig. 8(a) for the cases $H/L = 1/4$, $1/3$, and $5/12$. For comparison, the results of the droplet with $R_0 = 45$ are shown in Fig. 8(b). In the two figures a similar trend can be observed about the average moving velocity of the droplet when $H/L$ increases from $1/4$ to $5/12$, namely the maximum moving velocity of the droplet generally appears in the case $H/L = 1/3$. Meanwhile, from Fig. 8 we can also find that in the early stage the results of the cases $H/L = 1/3$ and $5/12$ are nearly the same. This is mainly because in the early stage the gravity has an important influence and the vapor flow below the droplet has not dominated the motion of the droplet.

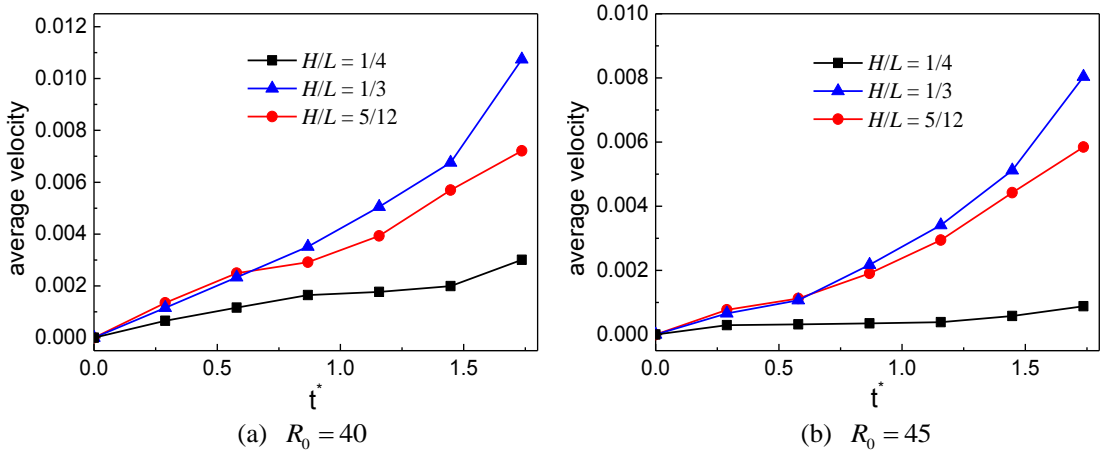

(a) $R_0 = 40$      (b) $R_0 = 45$

**Fig. 8** The average moving velocities of Leidenfrost droplets on the ratchet surfaces with $H/L = 1/4$, $1/3$, and $5/12$. The initial droplet radius is $R_0 = 40$ (left) and $R_0 = 45$ (right).

Furthermore, from each panel of Fig. 7 we can observe that the droplet gets smaller (i.e., the droplet evaporates more rapidly) when $H/L$ increases. Such a phenomenon is related to the following two changes. First, it can be found that the area of the heating surface is increased when $H/L$ increases. Second, with the increase of $H/L$, the vapor flow beneath the droplet will have more space to escape, which can be seen in Fig. 7. As a consequence, the thickness of the vapor layer between the ratchet peak and the bottom surface of the droplet will be reduced. Then the heat transfer between them, which is an important part of the whole heat transfer [9], will be enhanced.

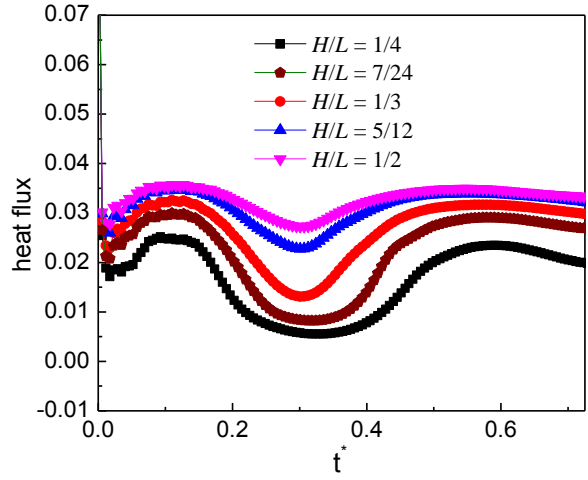

**Fig. 9** The transient heat flux at the initial ratchet peak ($R_0 = 40$).

To illustrate the above point, the heat flux at the initial ratchet peak (the peak covered by the droplets at the initial time, see Fig. 4(a)) is evaluated. Usually, the heat flux at a corner can be calculated according to the energy balance around the corner (see the textbooks of heat transfer, e.g., [58]). Here the phase change is not considered in calculating the heat flux with the energy balance treatment since in the present study the rate of the liquid-vapor phase change is a computational output not an *a priori* given quantity. The results are presented in Fig. 9 for the time $0 \leq t^* \leq 0.725$, during which the droplets are moving on the initial ratchet peak. From the figure we can clearly see that the heat flux increases when $H$ varies from $1/4$ to $1/2$. It is obvious that the heat flux of the case $H/L=1/4$ is much lower than that of the other cases and it can be found that the cases $H/L=5/12$ and $1/2$ give very close results. Previously, it has been shown that, among the five cases, the case $H/L=1/3$ is better than the others in terms of the moving velocity of the droplet. From Fig. 9 it can be seen that the heat flux given by this case is smaller than that of the cases $H/L=5/12$ and $1/2$. Nevertheless, it does not mean the case $H/L=1/3$ is inferior to these two cases when a high heat flux is required (e.g., for cooling systems), because within a given time we can place more droplets on the ratchet surface that leads to a large droplet velocity.

3.2.4 The Leidenfrost droplets on inclined ratchet surfaces

In this subsection we present some results about the process that a Leidenfrost droplet climbs uphill on a tilted ratchet surface. To date, there have been no numerical simulations about such a process and the first experimental study was conducted by Melling [57]. The numerical setup and the grid system are the same as those used in the above simulations except that the angle between the gravitational force and the dash-dotted line in Fig. 3 is no longer 90 degrees. The numerical results of the droplets with $R_0=40$ and $35$ on an inclined ratchet surface are displayed in Fig. 10 ($H/L=1/3$). The surface is inclined at an angle of 2 degrees to the horizontal. From the figure the climbing processes of the Leidenfrost droplets can be clearly observed. Similar to the results on horizontal ratchet surfaces, the results on tilted ratchet surfaces also show that the droplet with $R_0=35$ moves faster than the droplet with $R_0=40$, which indicates that the uphill acceleration is relatively large in the former case. Moreover, by comparing the results of the droplet with $R_0=40$ on horizontal and inclined ratchet surfaces (in Figs. 4 and 10), we can find that the moving velocity of the droplet is reduced on the inclined ratchet surface. Such a reduction of the droplet velocity is expected since the uphill acceleration provided by the vapor flow to the Leidenfrost droplet should overcome the downhill acceleration, which results from the gravitational force.

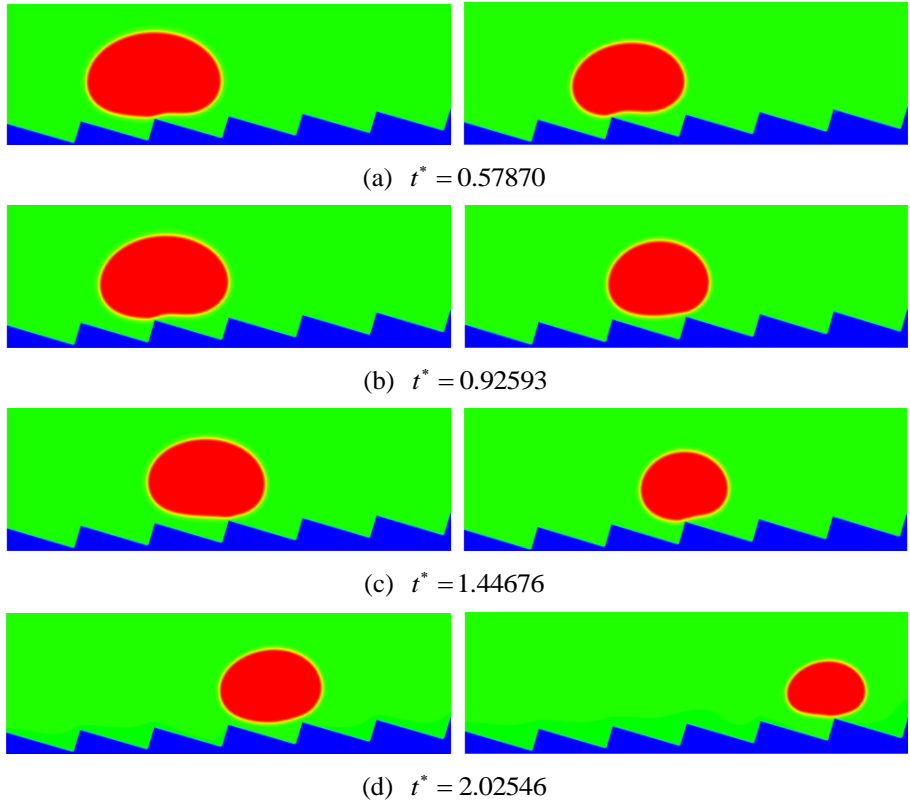

(a) $t^* = 0.57870$

(b) $t^* = 0.92593$

(c) $t^* = 1.44676$

(d) $t^* = 2.02546$

**Fig. 10** Snapshots of the motion of Leidenfrost droplets on a ratchet surface inclined at an angle of 2 degrees to the horizontal. The initial droplet radius is $R_0 = 40$ (left) and $R_0 = 35$ (right).

As the inclination angle of the ratchet surface increases, the influence of the downhill acceleration caused by the gravitational force will increase. To illustrate this point, the results of the droplet with $R_0 = 35$ on a ratchet surface inclined at an angle of 4 degrees to the horizontal are presented in Fig. 11. By comparing the results in Fig. 11 with the results in the right panel of Fig. 10, we can see that, with the increase of the inclination angle of the ratchet surface, the droplet moves downhill in the early stage owing to the downhill acceleration. Nevertheless, for the case in Fig. 11, with time going on, the vapor flow beneath the droplet is still able to support the climbing uphill process of the droplet, which can be seen from Fig. 11(d). The corresponding streamlines at $t^* = 1.15741$ and $1.44676$ are shown in Fig. 12. The variations of the vapor flow beneath the droplet can be clearly observed by comparing Fig.

12(b) with Fig. 12(a). At $t^* = 1.44676$, it can be found that, because of the collision between the droplet and the ratchet surface, the vapor flow beneath the droplet has changed its direction in the region that is located on the left side of the central ratchet peak, which is the reason why the droplet turns around.

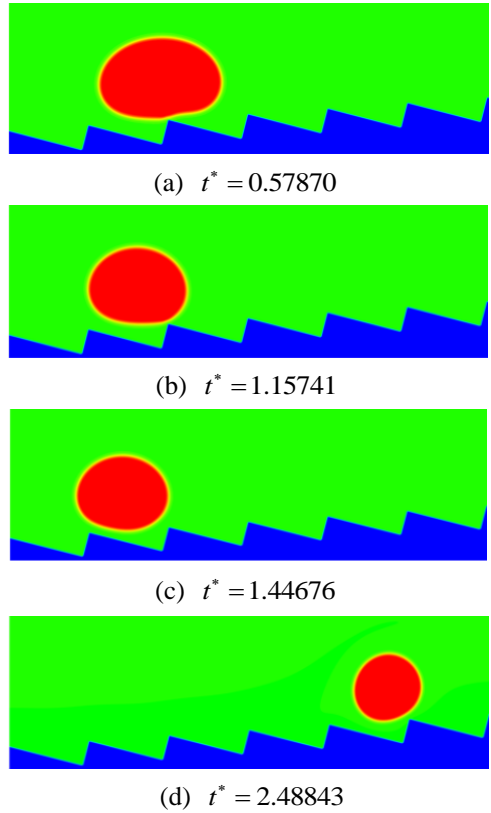

(a) $t^* = 0.57870$

(b) $t^* = 1.15741$

(c) $t^* = 1.44676$

(d) $t^* = 2.48843$

**Fig. 11** Snapshots of the motion of a Leidenfrost droplet on a ratchet surface inclined at an angle of 4 degrees to the horizontal ($H/L = 1/3$). The initial droplet radius is $R_0 = 35$.

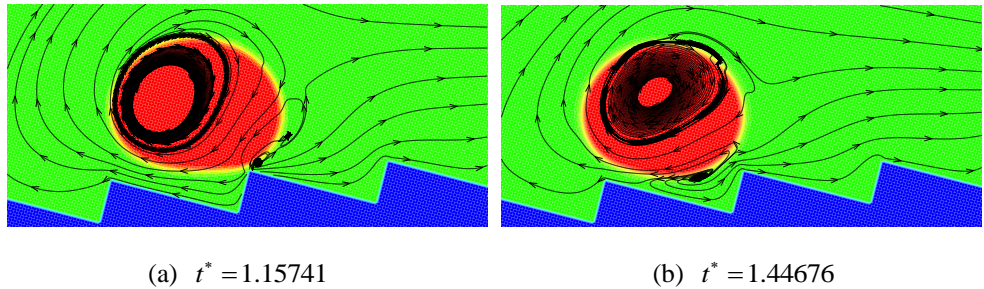

(a) $t^* = 1.15741$   (b) $t^* = 1.44676$

**Fig. 12** The streamlines at $t^* = 1.15741$ and $1.44676$.

When the inclination angle is further increased, it is believed that the downhill acceleration will gradually dominate the whole motion of the droplet. Different inclination angles have been investigated for the droplets with $R_0 = 35$, $40$, and $45$ on the ratchet surface with $H/L = 1/3$. The results are given in Fig. 13, where the filled circles denote the cases in which the Leidenfrost droplets can successfully climb uphill. If the droplet moves downhill in the early stage, the requirement of "a successful case" is that the downhill distance traveled by the droplet should be smaller than $\sqrt{L^2 + H^2}$. From Fig. 13 we can see that the maximum inclination angle at which a Leidenfrost droplet can still climb uphill successfully is different for the cases with different initial radii. To be specific, it can be seen that the maximum inclination angle decreases when the initial droplet radius increases, which is attributed to the reduction of the uphill acceleration from $R_0 = 35$ to $R_0 = 45$. For different droplets, the downhill acceleration caused by the gravity is the same when the inclination angle is given. However, the uphill acceleration is different, which can be seen clearly in Fig. 10.

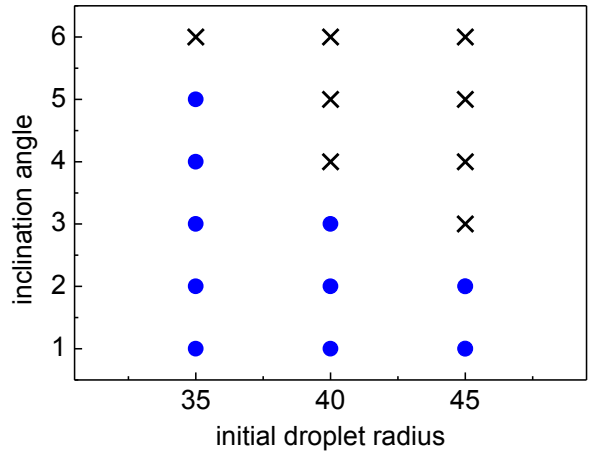

**Fig. 13** The achievable inclination angles (degree) regarding the climbing uphill processes of Leidenfrost droplets with different initial radii.

## 4. Summary and conclusions

In this work, an attempt has been made in investigating self-propelled Leidenfrost droplets on ratchet surfaces with numerical simulations. The numerical model is based on the lattice Boltzmann method, which consists of a pseudopotential multiphase lattice Boltzmann model for simulating the density and velocity fields and a finite-difference solver for the temperature field. The liquid-vapor phase change is driven by the temperature field via a non-ideal equation of state. The capability of the model for simulating evaporation has been validated through reproducing the well-known $D^2$ law.

The dynamic behavior of Leidenfrost droplets on horizontal ratchet surfaces has been investigated. Numerical results show that the self-propelled motion of Leidenfrost droplets originates from the asymmetry of the ratchets and the vapor flows below the droplets. It is found that the Leidenfrost droplets move in the direction toward the slowly inclined side from the ratchet peaks, which agrees with the direction of droplet motion observed in Linke *et al.*'s experiments. The effects of the initial droplet radius $R_0$ and the influences of the ratchet aspect ratio $H/L$ have been studied. For the considered ratchet surfaces (see Fig. 3), it has been found that there exists a critical value of $H/L$. Numerical results show that the droplet velocity increases when $H/L$ increases from $1/4$ to $1/3$. However, it decreases when $H/L$ is further increased. Moreover, we have also studied the performances of Leidenfrost droplets on inclined ratchet surfaces. Different inclination angles have been investigated. For the case of the droplet with $R_0 = 35$ on a ratchet surface inclined at an angle of 4 degrees to the horizontal, it is found that the droplet moves downhill in the early stage due to the downhill acceleration caused by the gravitational force. Later, the droplet turns around at a certain time with the help of the uphill acceleration, which is generated by the vapor flow beneath the droplet. The maximum inclination angle at which a Leidenfrost droplet can still climb uphill successfully is found to be related to the initial radius of the droplet.

In summary, we have numerically revealed some basic features of self-propelled Leidenfrost droplets on both horizontal and tilted ratchet surfaces. We hope the present work will stimulate more numerical studies of self-propelled Leidenfrost droplets from various aspects. As previously mentioned, the treatment of using a non-ideal equation of state for simulating liquid-vapor phase change is currently applicable to low or moderate density ratios. In the future, attention will be paid to the improvement of the model for simulating liquid-vapor phase change at large density ratios ($\sim 1000$), so as to enable quantitative comparisons with experimental studies. In addition, three-dimensional modeling will be considered in the future work, which may provide more information about the features of self-propelled Leidenfrost droplets.

**Acknowledgments**

This work was supported by the Los Alamos National Laboratory's Lab Directed Research & Development Program, the National Natural Science Foundation of China (No. 51506227), and the Foundation for the Author of National Excellent Doctoral Dissertation of China (No. 201439). Q. K. also acknowledges the support from a DOE NETL Unconventional Oil & Gas Project.